\documentclass[]{pasj02} 
\usepackage[switch,mathlines]{lineno} 

\jyear{2026}
\Received{}
\Accepted{}

\begin{document} 

\title{Molecular absorption lines toward the Galactic ISM detected at 89 GHz with VLBI}

\author{
Keisuke \textsc{Nakashima},\altaffilmark{1,2}\altemailmark\orcid{0000-0002-8764-9595} \email{k8024878@kadai.jp} 
Hiroshi \textsc{Imai},\altaffilmark{1,3,4}\altemailmark\orcid{0000-0002-0880-0091} \email{hiroimai@km.kagoshima-u.ac.jp}
and 
Youngjoo \textsc{Yun}\altaffilmark{5}\altemailmark\orcid{0000-0002-0822-2973} \email{yjyun@kasi.re.kr}
}
\altaffiltext{1}{Graduate School of Science and Engineering, Kagoshima University, 1-21-35 Korimoto, Kagoshima 890-0065, Japan}
\altaffiltext{2}{Nobeyama Radio Observatory, National Astronomical Observatory of Japan (NAOJ), National Institutes of Natural Sciences (NINS), 462-2, Nobeyama, Minamimaki, Minamisaku, Nagano 384-1305, Japan}
\altaffiltext{3}{Amanogawa Galaxy Astronomy Research Center, Graduate School of Science and Engineering, Kagoshima University, 1-21-35 Korimoto, Kagoshima 890-0065, Japan}
\altaffiltext{4}{Center for General Education, Institute for Comprehensive Education, Kagoshima University, 1-21-30 Korimoto, Kagoshima 890-0065, Japan}
\altaffiltext{5}{Korea Astronomy and Space Science Institute, 776 Daedukdae-ro, Yuseong-gu, Daejeon 34055, Korea}




\maketitle

\begin{abstract} 
As the first step to explore the Galactic foreground interstellar medium (ISM) with high angular resolution, we investigated the detectability of molecular absorption lines with very long baseline interferometry (VLBI). This paper reports our first observations of HCO$^+$ ${\it{J}}=1\rightarrow0$ and HCN ${\it{J}}=1\rightarrow0$ molecular absorption lines toward four extragalactic compact radio sources with the Korean VLBI Network (KVN). HCO$^+$ absorption lines were detected in all directions and HCN absorption lines in three directions. All of these detections, except for two HCO$^+$ detections, are reported here for the first time with VLBI. We find that the column density ratio $N(\mathrm{HCN})/N(\mathrm{HCO^+})$ obtained with VLBI is consistent with the trend established from observations with connected-element interferometers, demonstrating the feasibility of future VLBI studies of molecular absorption.
\end{abstract}


\begin{longtable}{ccccccc}
  \caption{Parameters of the background sources.}\label{tab:basis} \\
\hline\noalign{\vskip3pt}
  Source & Alias & \multicolumn{2}{c}{Galactic coordinates} & Observation dates & Continuum flux\footnotemark[$*$] & rms noise \\ [2pt]
    &  & $l$ [$\degree\:^\prime\:^{\prime\prime}$] & $b$ [$\degree\:^\prime\:^{\prime\prime}$] &  & $S_\mathrm{89 GHz}$ [Jy] & [mJy beam$^{-1}$] \\ [2pt]
\hline\noalign{\vskip3pt}
\endfirsthead
\hline\noalign{\vskip3pt}
  Source & Alias & \multicolumn{2}{c}{Galactic coordinates} & Observation dates & Continuum flux\footnotemark[$*$] & rms noise \\ [2pt]
    &  & $l$ [$\degree\:^\prime\:^{\prime\prime}$] & $b$ [$\degree\:^\prime\:^{\prime\prime}$] &  & $S_\mathrm{89 GHz}$ [Jy] & [mJy beam$^{-1}$] \\ [2pt]
\hline\noalign{\vskip3pt}
\endhead
\hline\noalign{\vskip3pt}
\endfoot
\hline\noalign{\vskip3pt}
\multicolumn{2}{@{}l@{}}{\hbox to0pt{\parbox{160mm}{\footnotesize
\hangindent6pt\noindent
\hbox to6pt{\footnotemark[$*$]\hss}\unskip%
  Variable, the values at the time of observation.
}\hss}}
\endlastfoot
  J0359+5057 & NRAO 150 & 150 22 38.22 & $-$01 36 13.46 & 2023 Nov 10 & 3.35 & 191 \\
  J0418+3801 & 3C 111 & 161 40 31.93 & $-$08 49 11.21 & 2023 Nov 11 & 0.69 & 48.6 \\
  J2202+4216 & BL Lac & 92 35 22.50  & $-$10 26 28.17 & 2023 Nov 14 & 9.47 & 470 \\
  J2253+1608 & 3C 454.3 & 86 06 39.85  & $-$38 11 01.74 & 2023 Nov 15 & 2.43 & 134 \\
\end{longtable}

\section{Introduction}\label{sec:1} 
The origin of structures found in molecular hydrogen clouds (typically $> 1$~pc) within the interstellar medium (ISM) remains a key open question in astrophysics. In particular, it is unclear whether these clouds originate from diffuse, structureless gas or from clusters of tiny-scale ($\ll 10^4$~au) structures that may eventually coalesce into seeds of much larger, dense molecular clouds. Tiny-scale structures have been established in the atomic phase (tiny-scale atomic structures, TSAS) and the ionized phase (tiny-scale ionized structures, TSIS) of the ISM (reviewed in \cite{Stan18}). In contrast, direct observational evidence for analogous structures in the molecular phase (tiny-scale molecular structures, TSMS; \cite{Andre04}) has remained elusive.\\
\indent Theoretically, thermal instability driven by collisions between warm neutral medium (WNM) or by the passage of shock waves is expected to produce cold neutral medium (CNM) \citep{Hennebelle}. Simulations further suggest that subsequent collisions between CNM fragments may lead to the formation of molecular structures on sub-parsec scales, down to ${\sim}0.05$~pc (${\sim}10^4$~au) \citep{InoInu}. In addition, the characteristic wavelength of thermal instability ranges from $\sim10^{-7}$ to $0.1$~pc \citep{Aota}, suggesting that molecular structures may form across a wide range of scales.\\
\indent Observationally, molecular absorption line surveys with connected-element interferometers toward compact background sources such as extragalactic quasars (QSOs) have detected a large number of molecular species (e.g., \cite{Ando16, Narita, LisztGerin25}). Absorption line observations are a powerful tool for exploring low-excitation gas that is largely inaccessible to molecular emission line observations. However, significant temporal variations in molecular absorption profiles have not been detected for up to 26~years \citep{Rybarczyk22}, and most are known to have highly stable profiles \citep{LiLu00}, leaving the existence of TSMS along individual lines of sight unconfirmed. Here note that H$_2$CO temporal fluctuations corresponding to structures of $\lesssim10$ au have been reported \citep{Marscher93, Moore&Marscher95}, but it has subsequently been suggested that they arise not from genuine density structures but rather from chemical or other inhomogeneities, or from fractal properties of the ISM \citep{Thoraval96, LiLu00, Araya14}. Direct observational evidence for TSMS has, however, recently emerged. \cite{Goldsmith25} used JWST to directly image heated TSMS at $140$--$350$~au via H$_2$ S(1) emission in the boundary region of the Taurus molecular cloud (visual extinction $A_V = 0.2$--$1.0$~mag). Nevertheless, whether comparable structures exist in cold, low-excitation gas remains an open question that absorption line observations are uniquely suited to address.\\
\indent Very long baseline interferometry (VLBI) provides milli-arcsecond (mas) angular resolution, making it in principle capable of absorption measurements toward compact background continuum sources. For spatially resolved background sources, such observations make it possible to probe the ISM with much narrower pencil beams than those accessible with connected arrays, and in the future may allow absorption to be measured toward individual continuum components of background sources. However, given the extremely limited number of Galactic molecular absorption detections with VLBI to date \citep{Han}, it remains unclear whether VLBI observations can reproduce molecular absorption features consistent with those observed with connected-element interferometers. Demonstrating this detectability is therefore a prerequisite for future VLBI studies of molecular absorption.\\
\indent In this paper, we present observations of HCO$^+$ and HCN absorption lines with the Korean VLBI Network (KVN) toward four QSOs for which high optical depths were previously confirmed in connected-array observations \citep{Ando16}. These VLBI observations detected HCO$^+$ absorption lines in two new directions, in addition to the two previously reported by \citet{Han}, and HCN absorption for the first time in three directions with an angular resolution of $\sim2$ mas. Section \ref{sec:2} describes the observations and data reduction procedures, and section \ref{sec:3} presents the results and discussion, including the column density ratios of $N(\mathrm{HCN})/N(\mathrm{HCO^+})$ on VLBI angular scales and the host Galactic structure for absorbing molecular gas. Details on possible temporal variations in HCO$^+$ and HCN absorption lines over a one-year period, based on KVN observations with even higher spectral resolution, will be reported in a forthcoming paper.

\section{Observations and data reduction}\label{sec:2} 
Our VLBI observations were conducted using three KVN stations (Yonsei, Tamna, and Ulsan) for the HCO$^+$ ${\it{J}}=1\rightarrow0$ (89.1885247 GHz), and the three hyperfine lines of HCN ${\it{J}}=1\rightarrow0$, ${\it{F}}=1\rightarrow1$ (88.6304156 GHz), ${\it{F}}=2\rightarrow1$ (88.6318475 GHz) and ${\it{F}}=0\rightarrow1$ (88.6339357 GHz) as absorption lines toward J0359+5057, J0418+3801, J2202+4216, and J2253+1608 in November 2023. Each source was observed for typically 10 hours in dual circular polarization. The observed bright sources were used as fringe finders and bandpass and phase calibrators as well. The exception was J0418+3801, which would be weak in the observed frequency band. Alternatively,  J0422+5324 was observed as a fringe finder and a bandpass calibrator although this source is out of our scientific data analysis due to too short exposure time. The system noise temperature was typically $\sim200$ K. Table \ref{tab:basis} gives the position, continuum flux and root-mean-square (rms) noise value of the obtained spectrum for each source.\\
\indent Received radio frequency signals from all the stations were sampled and filtered into two baseband channels (BBCs) each with a bandwidth of 512 MHz and then recorded at a rate of 8 Gbits s$^{-1}$ on a Mark 6 system. The recorded data were processed with the DiFX software correlator \citep{Deller}. The resulting data have 1024 spectral channels per BBC, with a spectral channel spacing of 500 kHz, corresponding to a velocity channel spacing of $\sim$1.69 km\,s$^{-1}$, although it will be improved in forthcoming observations, to a velocity resolution of $\sim$0.11 km\,s$^{-1}$ with 31.25 kHz per channel.\\
\indent Post-correlation processing was made using the NRAO AIPS (Astronomical Image Processing System) software \citep{Greisen} following standard imaging procedures. The correlated datasets were loaded, and each visibility amplitude was calibrated using the information of system noise temperature and antenna gain. To remove residual group-delays and fringe phases as well as temporal phase drifts, fringe fitting was performed using the background QSOs as calibrators. Subsequently, phase self-calibration was applied to the target source to correct for rapid fluctuations of residual phases. Here note that in the Tamna station, an oven-controlled crystal oscillator (OCXO) was used as a time and frequency standard instead of a hydrogen maser one commonly used at other stations. Given the short coherence time of signals from the OCXO (the fringe phase typically rotates by $2\pi$ radian within $\sim5$ seconds), its use in VLBI scientific observations is unusual. Nevertheless, the phase calibration solutions were obtained and applied at intervals of 3.2768 second. After that, the bandpass characteristics were calibrated using the cross-power spectra with a solution interval of 60 minutes. Inverse Fourier transforms were performed on the calibrated visibilities for image synthesis in which data weighting was applied with robustness set to 0 ($-4$ for uniform, $4$ for natural weighting). Note that for J2253+1608, we adopted a setting with robustness set to 4 to avoid artificially small CLEAN restoring beam. In all cases, the background sources are seen as almost point-like sources in the resulting images. The intensity profiles were normalized by each continuum flux level.

\begin{figure*}[htbp]
  \begin{center}
      \includegraphics[width=\textwidth]{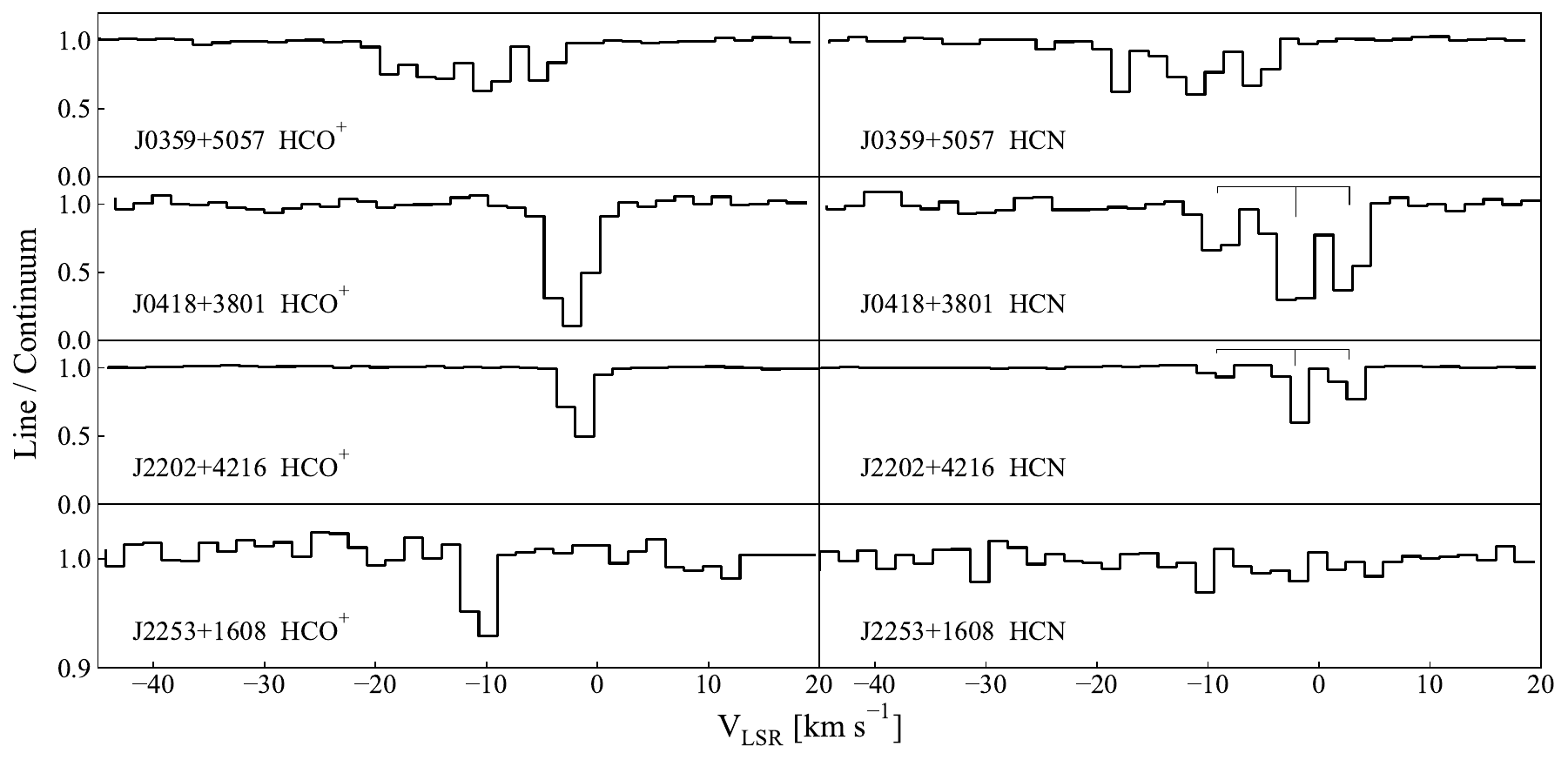}
  \end{center}
  \caption{Spectra of molecular absorption lines toward four background sources. The flux density scale of the absorption line profile was normalized with respect to that of the background emission.
  {Alt text: This figure consists of a four-row by two-column arrangement of absorption line spectra. From top to bottom, the rows correspond to the sources J0359+5057, J0418+3801, J2202+4216, and J2253+1608. The left column displays the HCO$^+$ ${\it{J}}=1\rightarrow0$ profiles, and the right column displays the HCN ${\it{J}}=1\rightarrow0$, ${\it{F}}=1\rightarrow1, 2\rightarrow1, 0\rightarrow1$ profiles. The vertical axis represents the line-to-continuum flux ratio with values spanning approximately zero to one, and the horizontal axis represents the velocity in the local standard of rest frame in kilometer per second, covering minus fifty to plus twenty kilometer per second.}}
  \label{fig:combined_spectra_graph}
\end{figure*}

\begin{longtable}{cccc}
  \caption{Parameters derived from HCO$^+$ absorption components exceeding $5\sigma$.}\label{tab:HCO+_parm} \\
\hline\noalign{\vskip3pt}
  Source & $V_{\mathrm{LSR}}$ range & Integrated optical depth & Column density \\ [2pt]
    & [km\,s$^{-1}$] & $\sum{\tau dv}$ [km\,s$^{-1}$] & $N_{\mathrm{total}}^{\mathrm{HCO^+}}$ [$10^{12}$ cm$^{-2}$] \\ [2pt]
\endfirsthead
\hline\noalign{\vskip3pt}
  Source & $V_{\mathrm{LSR}}$ range & Integrated optical depth & Column density \\ [2pt]
    & [km\,s$^{-1}$] & $\sum{\tau dv}$ [km\,s$^{-1}$] & $N_{\mathrm{total}}^{\mathrm{HCO^+}}$ [$10^{12}$ cm$^{-2}$] \\ [2pt]
\hline\noalign{\vskip3pt}
\endhead
\hline\noalign{\vskip3pt}
\endfoot
\hline\noalign{\vskip3pt}
  J0359+5057 & [$-19.6,\ -2.8$] & $4.48\pm{-0.30}$ & $4.97_{-0.33}^{+0.34}$ \\
  J0418+3801 & [$-4.8,\ 0.2$] & $6.94_{-0.70}^{+0.88}$ & $7.70_{-0.78}^{+0.98}$ \\
  J2202+4216 & [$-3.7,\ 1.4$] & $1.84\pm{0.07}$ & $2.04\pm{0.08}$ \\
  J2253+1608 & [$-10.7,\ -9.0$] & $0.13\pm{-0.02}$ & $0.14\pm{0.02}$ \\
\end{longtable}

\begin{longtable}{cccc}
  \caption{Parameters derived from HCN absorption components exceeding $5\sigma$.}\label{tab:HCN_parm} \\
\hline\noalign{\vskip3pt}
  Source & $V_{\mathrm{LSR}}$ range & Integrated optical depth & Column density \\ [2pt]
    & [km\,s$^{-1}$] & $\sum{\tau dv}$ [km\,s$^{-1}$] & $N_{\mathrm{total}}^{\mathrm{HCN}}$ [$10^{12}$ cm$^{-2}$] \\ [2pt]
\endfirsthead
\hline\noalign{\vskip3pt}
  Source & $V_{\mathrm{LSR}}$ range & Integrated optical depth & Column density \\ [2pt]
    & [km\,s$^{-1}$] & $\sum{\tau dv}$ [km\,s$^{-1}$] & $N_{\mathrm{total}}^{\mathrm{HCN}}$ [$10^{12}$ cm$^{-2}$] \\ [2pt]
\hline\noalign{\vskip3pt}
\endhead
\hline\noalign{\vskip3pt}
\endfoot
\hline\noalign{\vskip3pt}
  J0359+5057 & [$-18.7,\ -3.5$] & $4.22\pm{0.31}$ & $8.07_{-0.58}^{+0.59}$ \\
  J0418+3801 & [$-10.5,\ 4.7$] & $8.93_{-1.02}^{+1.12}$ & $17.06_{-1.95}^{+2.14}$ \\
  J2202+4216 & [$-11.0,\ 4.2$] & $1.80\pm{0.07}$ & $3.44\pm{0.13}$ \\
  J2253+1608 & [$-11.1,\ -9.4$] & $<0.05\pm{0.01}$ & $<0.10\pm{0.02}$ \\
\end{longtable}

\begin{figure}
  \begin{center}
    \includegraphics[width=0.94\columnwidth]{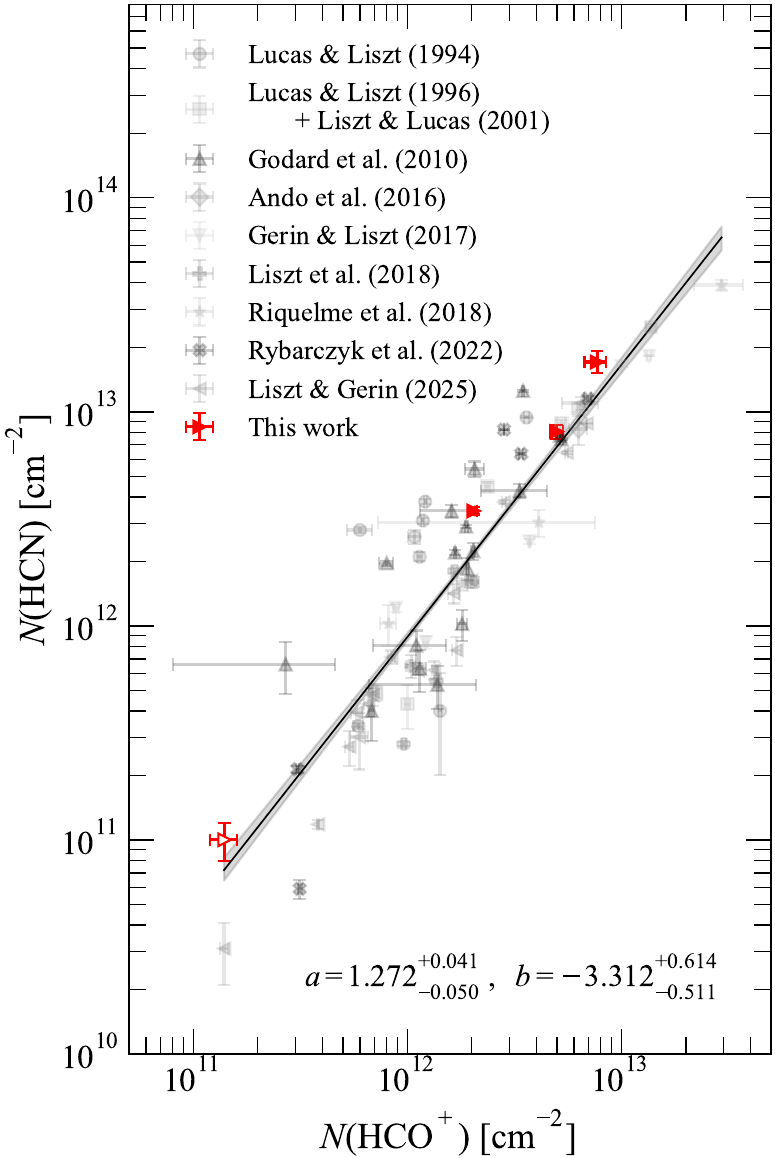}
  \end{center}
  \caption{Comparison between the column densities of HCO$^+$ and HCN obtained from the detected absorption components. The data from this study and those previously derived from Galactic diffuse gas (see the main text for details) are identified in the legend. A Monte Carlo error-propagation method using split-normal sampling was applied to incorporate asymmetric uncertainties in both axes for deriving a linear fit to the column density relation. The resulting relation, $\log{N(\mathrm{HCN})}=a\log{N(\mathrm{HCO^+})}+b$, is shown by a solid line, with a shaded region indicating the $1\sigma$ confidence interval. The fitted values of $a$ and $b$ are displayed in the lower-right corner of the panel.
  {Alt text: This figure presents a scatter plot comparing the logarithmic column densities of HCO+ and HCN, expressed in square centimeters. The four data points obtained in this study represent column densities integrated along individual lines of sight. The remaining points are derived from molecular absorption measurements toward forty-three lines of sight collected from multiple earlier publications. A linear relation between the two column densities is plotted together with its one-sigma confidence interval, and the numerical values of the fitted slope and intercept appear within the plotting area.}}
  \label{fig:N_ratio}
\end{figure}

\section{Results and discussion}\label{sec:3} 
\subsection{Profiles of the molecular absorption lines}\label{sec:3-1}
We have successfully detected HCO$^+$ absorption lines in all four directions and HCN lines in three directions, except J2253+1608. Figure \ref{fig:combined_spectra_graph} shows their spectra. In observations with the mas-scale VLBI beam, although HCO$^+$ absorption lines toward J0359+5057 and J2202+4216 were first reported by \citet{Han}, the remaining detections are presented here for the first time. The HCN spectra are plotted with the ${\it{F}}=2\rightarrow1$ component as a velocity reference. The intrinsic strength ratios, 1:5:3 from \citet{Goicoechea}, and velocity spacings, calculated from the rest-frequency intervals, of the hyperfine components are shown in the top row from left to right, except for J0359+5057, where the hyperfine components are blended within the current velocity resolution and J2253+1608, whose HCN spectrum shows no significant detection.\\
\indent Tables \ref{tab:HCO+_parm} and \ref{tab:HCN_parm} summarize the local-standard-of-rest (LSR) velocity ranges, integrated optical depths, and column densities for the detected HCO$^+$ and HCN absorption features, respectively. The LSR velocity ranges are identified from components exceeding $5\sigma$ of the spectral rms noise. Toward J2253+1608, no HCN absorption exceeding $5\sigma$ was detected. We therefore report an upper limit for the integrated optical depth and column density.\\
\indent The optical depth $\tau$ was derived using equations (1) and (2), which are identical to equations (A1) and (A2) of \citet{Ando16}, originally derived from \citet{Greaves}, for each bin of the extracted velocity components as follows:
\begin{align}
  \tau &= -\ln\left[1-\frac{T_\mathrm{line}}{J(T_{\mathrm{ex}})-J(T_{\mathrm{CMB}})+T_{\mathrm{cont}}}\right],\\
  &\text{where}\quad J(T) = \frac{h\nu}{k}\frac{1}{\exp\left(\frac{h\nu}{kT}\right)-1}.
\end{align}
Here $T_{\mathrm{line}}$ is the main beam brightness temperature of the line, which takes negative values for absorption lines, $T_{\mathrm{CMB}} = 2.73$ K the cosmic microwave background (CMB) temperature, $T_\mathrm{ex}$ the excitation temperature that is assumed here to be equal to $T_{\mathrm{CMB}}$ \citep{Godard10, Luo20}, and $T_{\mathrm{cont}}$ the continuum antenna temperature. The uncertainty in the optical depth was estimated by varying the normalized absorption depth by $1\sigma$ for each velocity bin and recalculating $\tau$ and $\sum{\tau dv}$.\\
\indent The total column density $N_{\mathrm{total}}\equiv\sum{Ndv}$ was derived based on equation (A3) of \citet{Ando16} as follows.
\begin{align}
  N_{\mathrm{total}}&=\frac{3h}{8\pi^3S\mu^2}\frac{Q(T_{\mathrm{ex}})\exp\left(\frac{E_l}{kT_{\mathrm{ex}}}\right)}{\left[1-\exp\left(-\frac{h\nu}{kT_{\mathrm{ex}}}\right)\right]}\sum{\tau dv}\nonumber\\
  &\equiv F(T_{\mathrm{ex}})\sum{\tau dv},
\end{align}
where $S$ is the intrinsic line strength, $\mu$ the permanent electric dipole moment of the molecule, $Q(T)$ the partition function, and $E_l$ the lower energy level of the transition. The conversion factors $F(T_{\mathrm{ex}}=2.73\ \mathrm{K})$ are set to $1.11\times10^{12}$ and $1.91\times10^{12}$ cm$^{-2}$\,km$^{-1}$\,s for HCO$^+$ and HCN, respectively. These values are listed in table 6 of \citet{Ando16}. The uncertainty in the column density was derived from that in the integrated optical depth.\\
\indent Figure \ref{fig:N_ratio} shows the plot of the column densities $N(\mathrm{HCN})$ and $N(\mathrm{HCO^+})$, with the four samples detected in this study indicated by red points. The gray data points present the values previously derived from 43 lines of sight toward which both lines were detected \citep{LuLi94, LuLi96, LiLu01, Godard10, Ando16, GerinLiszt17, Liszt18, Riquelme18, Rybarczyk22, LisztGerin25}, including directions in the inner and outer Galaxy, the Galactic bulge, and the outskirts of the Chamaeleon cloud complex. Note that these data points are plotted without duplication. We used a Monte Carlo error-propagation method based on split-normal sampling in log-log space in order to derive the linear fit to the column density relation and its  parameter uncertainties. A total of 50000 realizations were generated by perturbing every data point in the referenced dataset according to its individual asymmetric uncertainties, and a linear regression was performed for each perturbed dataset. The slope and intercept of the resulting relation were derived from the medians of their posterior-like distributions (solid line in figure \ref{fig:N_ratio}). The slope is found to be $1.272^{+0.041}_{-0.050}$ in this work, while previous studies reported the values of $1.97\pm0.43$ (weighted by column density) and $1.47\pm0.86$ (not weighted) in \citet{LiLu01}, $1.9\pm0.9$ in \citet{Godard10}, $1.96$ in \citet{Ando19}, $1.7\pm0.9$ in \citet{Rybarczyk22}, and $1.25\pm0.06$ in \citet{LisztGerin25}. The column density ratios obtained in this study roughly follow the trend derived in the previous data, indicating that VLBI observations successfully reproduce this relationship.

\subsection{Galactic structures hosting the absorption components}\label{sec:3-2} 
We compared the velocity ranges of the absorption lines detected in this study with those known from previous studies to identify the Galactic structures hosting the absorbing gas. The absorption lines toward J0359+5057 likely arise from widely distributed Galactic gas, consistent with the broad velocity range reported by \citet{LiLu00}, which includes several diffuse clouds mentioned in \citet{Liszt05} and \citet{Pety08}. The absorption lines toward J0418+3801 and J2202+4216 are likely to originate from the California molecular cloud \citep{LuLi98, Lada} and the Lacerta molecular cloud \citep{Rybarczyk22}, respectively.\\
\indent Finally, the absorption lines toward J2253+1608 have not been located with any specific molecular cloud in previous studies. Therefore, we estimated the distance using the Galactic 3D dust map \citep{Edenhofer2024} and the kinematic distance (KD) using the KD calculation tool \citep{Wenger}. We assumed that an HCO$^+$ absorption line without any hyperfine structure corresponds to a single gas structure, and used the peak values of the HCO$^+$ absorption lines as the measured LSR velocities for each distance calculation. For the former distance, we estimated the distance to the dust structure corresponding to the gas structure of CO and H\emissiontype{I} \citep{Dame22, HI4PI} in the same velocity range as the HCO$^+$. Note that this dust map does not trace beyond $1250$ pc from the Sun. The estimated distance using this method was $262.8\pm2.1$ or $397.9\pm2.5$ pc. For the latter distance, we first corrected the measured LSR velocities based on the values from \citet{Reid}, and then used the KD calculation tool \citep{Wenger}. This tool calculates the KD using the Monte Carlo method, based on the Galactic rotation curve and updated solar motion parameters derived in \citet{Reid}, and returns the KD uncertainties. The resulting estimated distance was $3650^{+850}_{-1250}$ pc.\\
\indent The large discrepancy between the two distance estimates may reflect the limitations of applying KDs to high Galactic latitude gas. If the absorbing gas is located at the dust-based distance ($\sim260$--$400$ pc), its vertical distance from the Galactic mid-plane would be $|z| \sim 160$--$247$ pc. This value is somewhat larger than the typical scale height of molecular gas layer in the Galactic disk, $\sim 120$ pc \citep{Heyer&Dame}, but still comparable within the same order of magnitude. In contrast, adopting the KD would place the gas at $|z| \sim 2260$ pc below the Galactic mid-plane, which is significantly larger than the reported thickness of both the molecular and atomic gas layers in the Milky Way \citep{Heyer&Dame, Nakanishi&Sofue}. Although the current dust map does not trace structures beyond $1250$ pc from the Sun and therefore cannot exclude a more distant origin, the dust-based distance is more consistent with the known vertical distribution of Galactic molecular gas. These considerations suggest that the absorbing gas toward J2253+1608 is likely associated with nearby diffuse structures rather than distant gas in the outer Galaxy.

\begin{ack}
We thank to the staff of the KVN who helped to operate the array and to correlate the data. The KVN is operated by the Korea Astronomy and Space Science Institute (KASI). The KVN observations and correlations are supported through the high-speed network connections among the KVN sites provided by the KREONET (Korea Research Environment Open NETwork), which is managed and operated by the Korea Institute of Science and Technology Information (KISTI). We also thank Dr. J. R. Dawson for fruitful discussions.
\end{ack}

\section*{Funding}
KN and HI were supported by JSPS KAKENHI 21H04524. KN also acknowledges support from the Yoshida Scholarship Foundation.

\end{document}